\documentclass{new_tlp}
\usepackage{mathptmx}
\usepackage{verbatim}
\usepackage{amssymb}  
\usepackage{epstopdf}
\usepackage{epsfig}
\usepackage{subfigure}
\usepackage[justification=centering]{caption}
\usepackage[noend]{algorithmic}
\usepackage{algorithm2e}
\usepackage{textcomp}
\usepackage{booktabs} 
\let\proof\relax

\usepackage{amsthm} 
\theoremstyle{definition}
\newtheorem{definition}{Definition}[section]
\let\olddefinition\definition
\renewcommand{\definition}{\olddefinition\itshape}
\newtheorem{example}{Example}[section]
\newcommand\bcmdtab{\noindent\bgroup\tabcolsep=0pt%
  \begin{tabular}{@{}p{10pc}@{}p{20pc}@{}}}
\newcommand\ecmdtab{\end{tabular}\egroup}

  \title[The P-Box CDF-Intervals]
        {The P-Box CDF-Intervals: Reliable Constraint Reasoning with Quantifiable Information}
  \author[A. SAAD, T. FR\"{U}HWIRTH and C. GERVET]
         {AYA SAAD, THOM FR\"{U}HWIRTH\\
         Universit\"{a}t Ulm, Germany
         \email{{ayas}@aucegypt.edu,{thom.fruehwirth}@uni-ulm.de}
         \and CARMEN GERVET\\
         Universit$\acute{e}$ de Savoie, France
         \email{{gervetec}@univ-savoie.fr}
         }
\begin{document}
\label{firstpage}
\maketitle
  \begin{abstract}
This paper introduces a new constraint domain for reasoning about data with uncertainty. It extends convex modeling with the notion of p-box to gain additional quantifiable information on the data whereabouts. Unlike existing approaches, the p-box envelops an unknown probability instead of approximating its representation. The p-box bounds are uniform cumulative distribution functions ({\em cdf}) in order to employ linear computations in the probabilistic domain. The reasoning by means of p-box {\em cdf}-intervals is an interval computation which is exerted on the real domain then it is projected onto the {\em cdf} domain. This operation conveys additional knowledge represented by the obtained probabilistic bounds. The empirical evaluation of our implementation shows that, with minimal overhead, the output solution set realizes a full enclosure of the data along with tighter bounds on its probabilistic distributions.
  \end{abstract}
\section{Introduction}
In this paper, we propose a novel constraint domain for reasoning about data with uncertainty. Our work was driven by the practical usage of reliable approaches in Constraint Programming (CP). These approaches tackle large scale constraint optimization (LSCO) problems associated with data uncertainty in an intuitive and tractable manner. Yet they have a lack of knowledge when the data whereabouts are to be considered. These whereabouts often indicate the data likelihood or chance of occurence, which in turn, can be ill-defined or have a fluctuating nature. It is important to know the source and type of the data whereabouts in order to reason about them. The purpose of our framework is to intuitively describe data coupled with uncertainty or following unknown distributions without losing any knowledge given in the problem definition. We extend the {\em cdf}-intervals approach \cite{saad2010constraint} with a p-box structure \cite{ferson2003constructing} to obtain a safe enclosure. This enclosure envelops the data along with its whereabouts with two distinct quantile values, each is located on a {\em cdf}-uniform distribution \cite{saadcdf}. 
This paper contains the following contributions: (1) a new uncertain data representation specified by p-box {\em cdf}-intervals, (2) a constraint reasoning framework that is used to prune variable domains in a p-box {\em cdf}-interval constraint relation to ensure their local consistency, (3) an experimental evaluation, using an inventory management problem, to support our argument by comparing the novel framework with existing approaches in terms of expressiveness and tractability. The expressiveness, in our comparison, measures the ability to model the uncertainty provided in the original problem, and the impact of this representation on the solution set realized. On the other hand, the tractability measures the system time performance and scalability. The experimental work shows how this novel domain representation yields more informed results, while remaining computationally effective and competitive with previous work. \vspace{-1\baselineskip}
\section{Preliminaries}
Models tackling uncertainty are classified under the set of plausibility measures \cite{halpern2003reasoning}. They are categorized as: possibilistic and probabilistic. Convex models, found in the world of {\em fuzzy} and interval/robust programming, are favored when ignorance takes place. They are adopted in the CP paradigm in {\em fuzzy} Constraint Satisfaction Problems (CSPs) \cite{dubois1996possibility}, soft CSPs \cite{bistarelli2002soft}, numerical CSPs \cite{benhamou1997applying} and uncertain CSPs (UCSPs) \cite{yorke2009}. Probabilistic models are best adopted when the data has a fluctuating nature. They are the heart of the probabilistic CP modeling found in valued CSP \cite{schiex1995}, semirings \cite{bistarelli1999}, stochastic CSPs \cite{walsh2008}, scenario-based CSPs \cite{tarim2006}, mixed CSPs \cite{fargier1996} and dynamic CSPs \cite{climent2014robustness}.
Techniques adopting convex modeling are characterized to be more conservative. They can often consider many unnecessary outcomes along with important ones. Due to this conservative property, operations exerted on convex models are tractable and scalable because they are exerted on the convex bounds only. On the other hand, probabilistic approaches add a quantitative information that expresses the likelihood, yet these approaches impose assumptions on the distribution shape in order to conceptually deal with it in a mathematical manner. Moreover, probabilistic mathematical computations are very expensive because they often depend on the non-linear probability shape. 

Our objective is to introduce a novel framework (the p-box {\em cdf}-intervals) that combines techniques from the convex models, to take advantage of their tractability, with approaches revealing quantifiable information from the probabilistic and stochastic world, to take advantage of their expressiveness. Our framework is based on CP concepts \cite{jaffar1987} because they proved to have a considerable flexibility in formulating real-world combinatorial problems. In the CP paradigm, we aim at building descriptive algebraic structures which are easily embedded into declarative programming languages. These structures are heavily used in the problem solving environment by specifying conditions that need to be satisfied and allow the solver to search for feasible solutions. Next we demonstrate how to intuively represent the uncertainty already given in the problem definition in order to reason about it by means of the p-box {\em cdf}-intervals. We also compare our novel representation of the data uncertainty with existing possibilistic and probabilistic approaches in order to demonstrate the model expressiveness. This representation is input to the solver with a new domain specification. We consequently define how to reason about this new specification and show how reasoning by means of p-box {\em cdf}-intervals proved to be tractable. Accordingly, we can claim that combining reasoning techniques from convex models with quantifiable information from probabilistic models yields a novel model that is together tractable and expressive. \vspace{-1\baselineskip}
\section{Input Data Representation}\label{sec:datarepresentation}
Quantifiable information is often available during the data collection process, but lost during the reasoning because it is not accounted for in the representation of the uncertain data. This information however is crucial to the reasoning process, and the lack of its interpretation yields erroneous reasoning because of its absence in the produced solution set. It is always necessary to quantify uncertainty that is naturally given in the problem definition in order to obtain robust and reliable solutions.
In this section, we show how to compute the confidence interval in the modeling approaches of the convex, possibilistic and probabilistic worlds, then we compare them with the input representation of the {\em cdf}-intervals and the p-box {\em cdf}-intervals. Given a data set of $n$ distinct values, the generic construction of the confidence possibilistic/probabilistic interval, in a measurement process of a population $m$, $m \neq n$, follows the steps below: 
\begin{enumerate}
\item Data is collected and $n$ quantiles (data values) are distinguished, each is represented by $x_i$. 
\item The probability distribution function ({\em pdf}) of the genuine observations is derived from $\frac{(x_i \mbox{Freq}_i)}{\sum^n_1 x_i \mbox{Freq}_i}$, where $\mbox{Freq}_i$ is the number of times $x_i$ is observed.
\item The average value of the observations, $\bar{x} = \frac{x_1 \mbox{Freq}_1 + \dots + x_n \mbox{Freq}_n}{\sum^n_1 x_i \mbox{Freq}_i}$ and their standard deviation, $\sigma = \sqrt{\frac{1}{n}\sum_1^n (x_i - \bar{x})^2}$ are computed.  
\item The probabilistic/ possibilistic distributions are derived from the average and the standard deviation values. Based on the \cite{gum1995guide} any probability distribution (parametric/non-parametric) is typically approximated to the nearest Normal distribution.
\item Computation and reasoning are based on the derived distributions since pointwise operations are computationally expensive.
\end{enumerate}
\begin{figure}[h]
\centering
\subfigure[]{\includegraphics[scale=0.16]{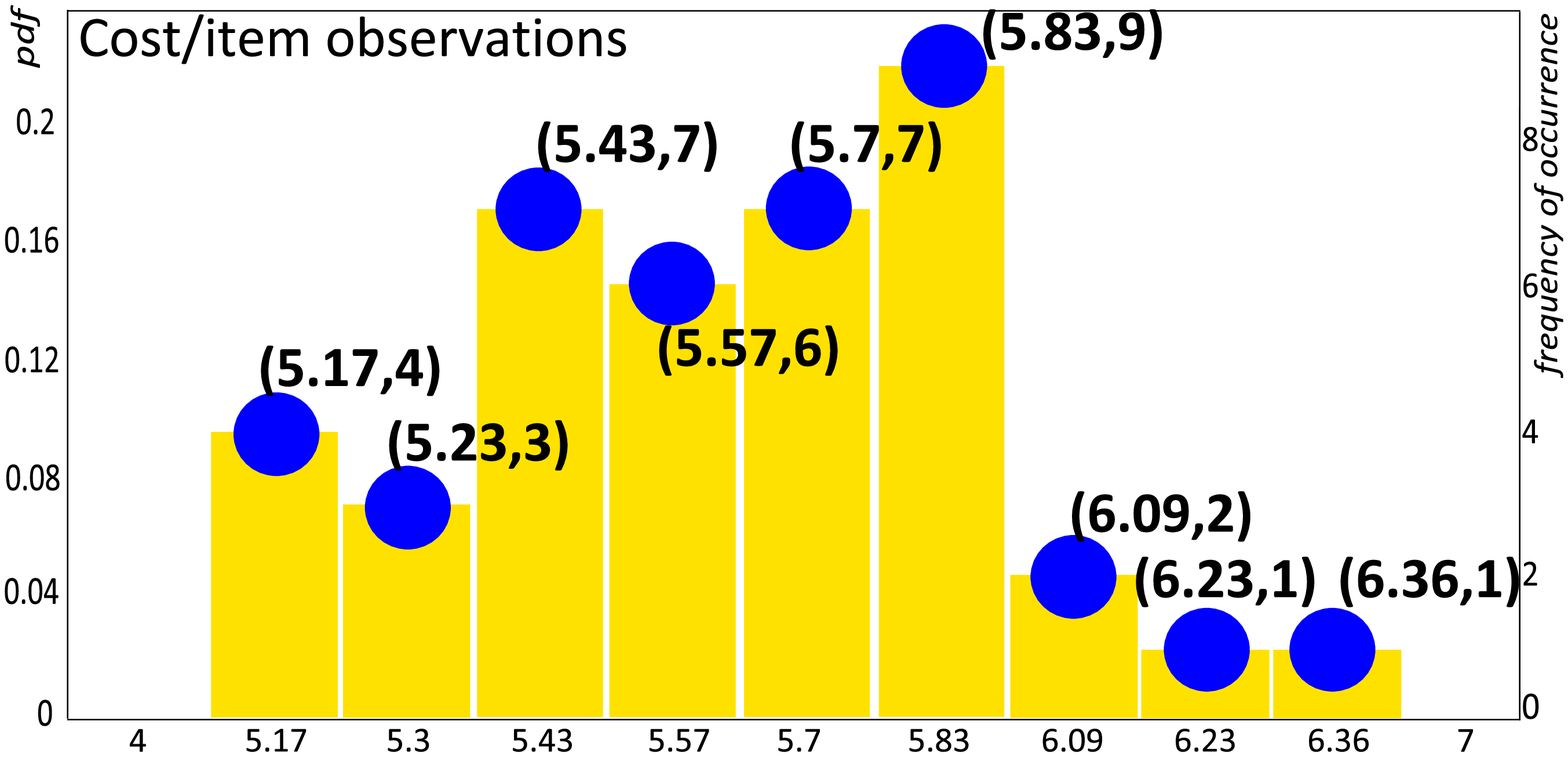}}
\subfigure[]{\includegraphics[scale=0.16]{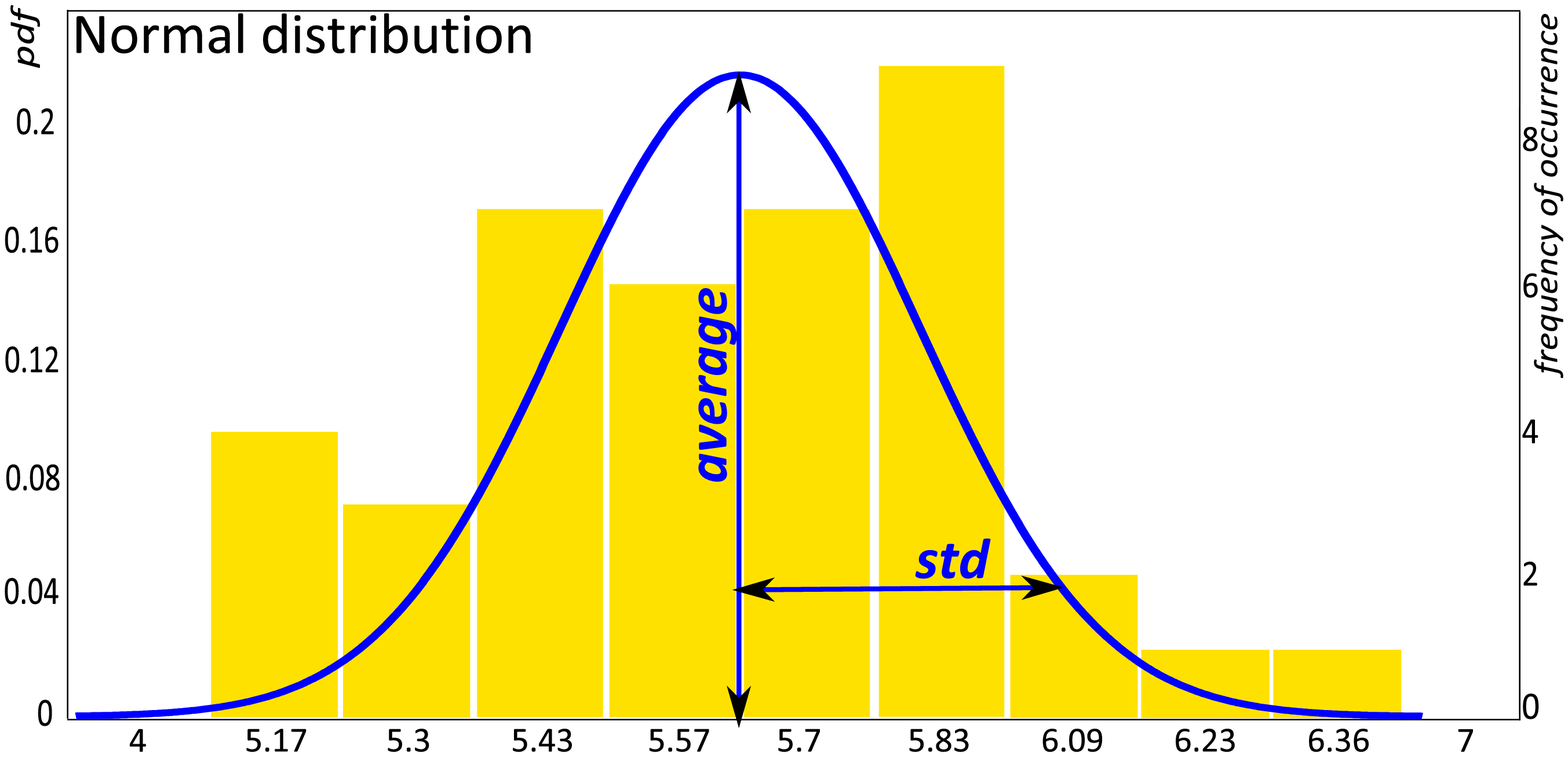}}
\subfigure[]{\includegraphics[scale=0.16]{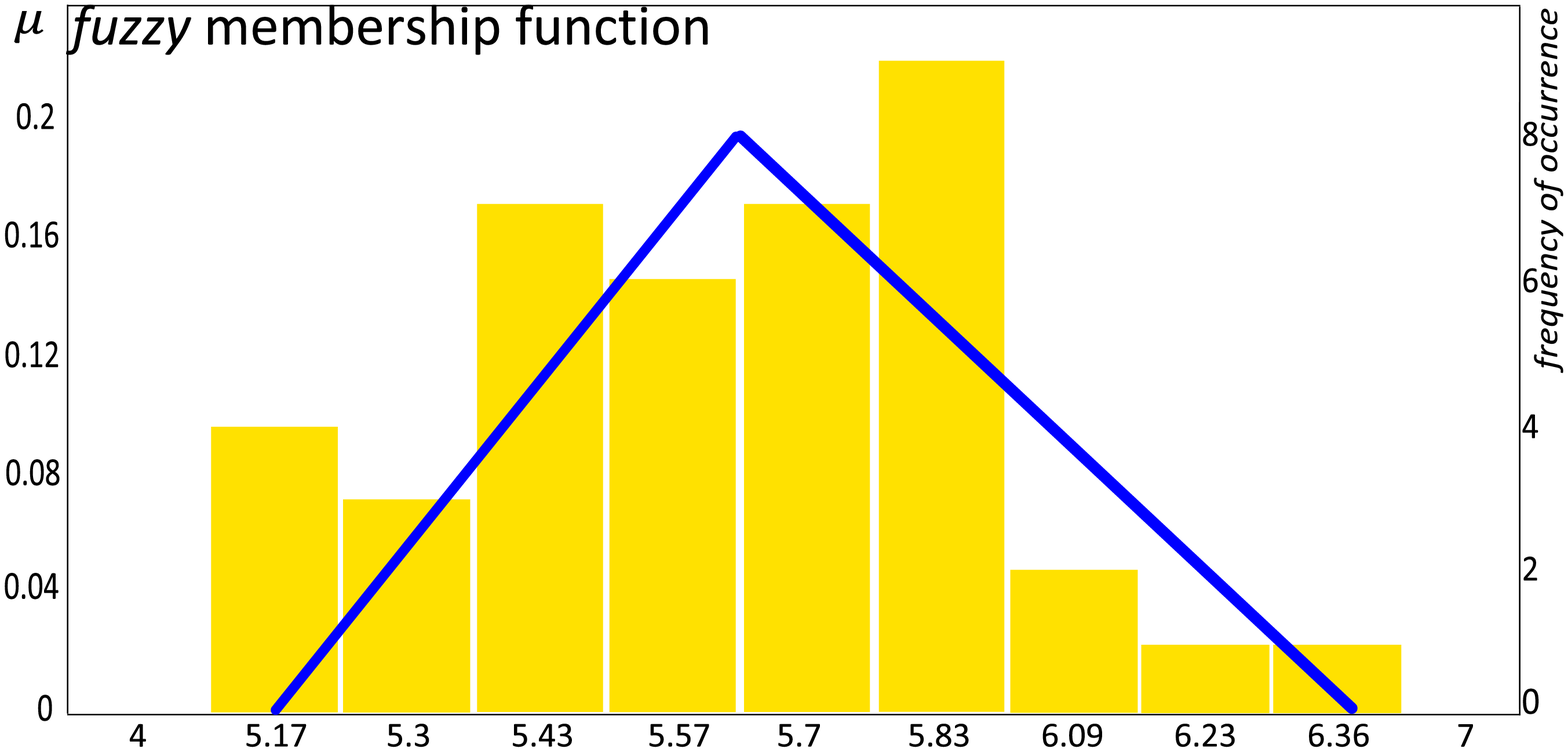}}
\caption{Varying cost of the steel stud item and its probability histogram: \\(a) genuine observations (b) Normal distribution (c) {\em fuzzy} distribution} \label{fig:costperitemobservationshistogram}
\end{figure}
\begin{example}\label{ex:costperitemdataobservation}
Consider, as a running example, the varying cost observations of a steel stud manufacturing item. Fig. \ref{fig:costperitemobservationshistogram}(a) illustrates the cost variations along with their corresponding frequencies of occurrence. For instance, the point $(5.17,4)$ is the amount of the cost/item, equal to $5.17$, and observed $4$ times during the past $2$ years (corresponding to a population $m = 40$). Nine is the number of distinct measured quantiles. The minimum and the maximum observed values, in this example, are $5.17$ and $6.36$ respectively. 
\end{example}
\paragraph{Computing the probabilistic/ possibilistic distribution.}\label{subsec:Computingthepdf}
The genuine {\em pdf} of the observed data, and its corresponding Normal distribution as well as its approximated possibilistic distribution are computed using the average and standard deviation. Recall from Example \ref{ex:costperitemdataobservation}, the point $(5.17,4)$ has a probability $\frac{(x_i \mbox{Freq}_i)}{\sum^n_1 x_i \mbox{Freq}_i} = 0.1$. The calculated average and standard deviation of the observed population are $5.6$ and $0.28$ respectively. From the two calculated values, we can derive the nearest Normal probability distribution and the {\em fuzzy} membership function as shown in Fig. \ref{fig:costperitemobservationshistogram} (b) and (c).
\begin{figure}[h]
\centering
\subfigure[]{\includegraphics[scale=0.16]{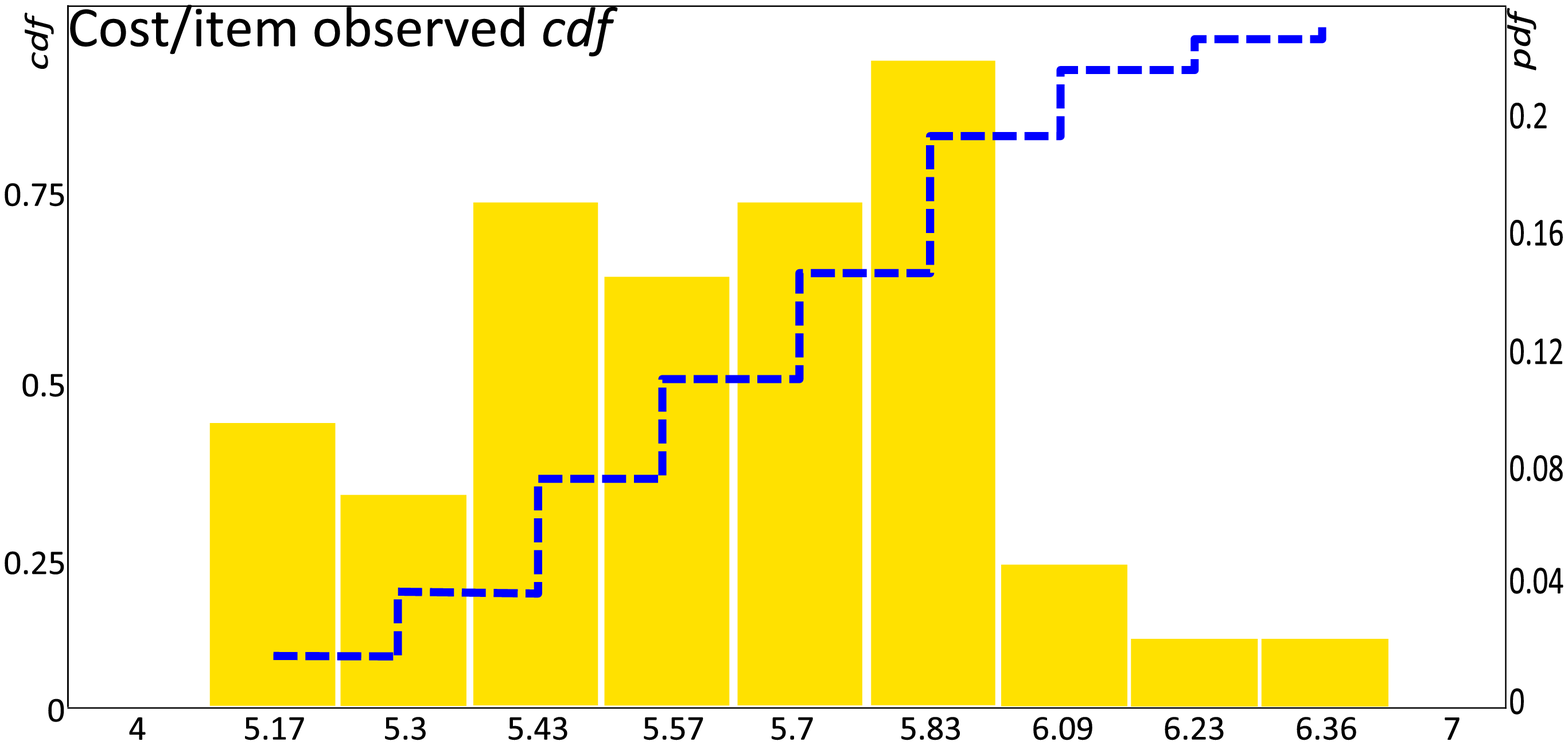}}
\subfigure[]{\includegraphics[scale=0.16]{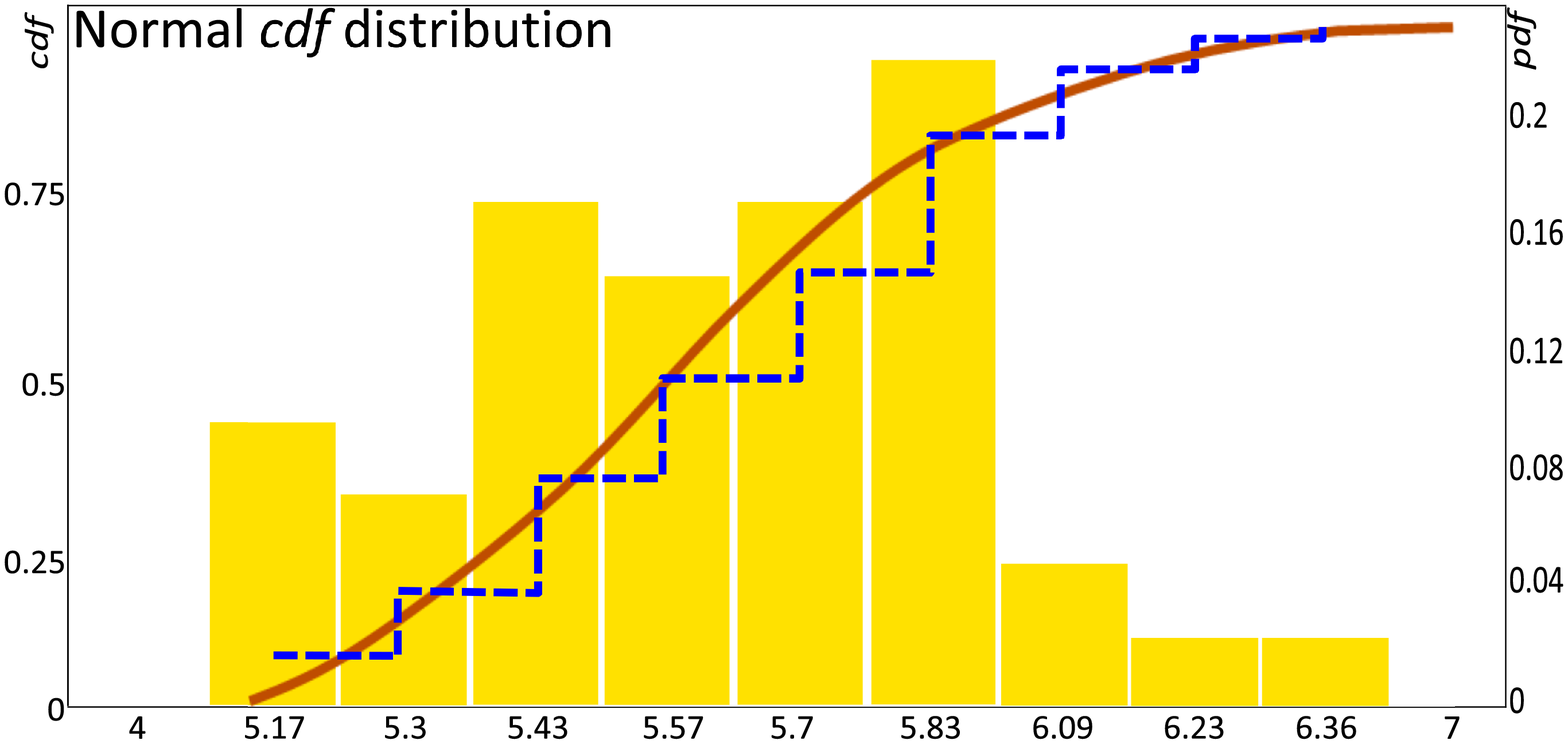}}
\subfigure[]{\includegraphics[scale=0.16]{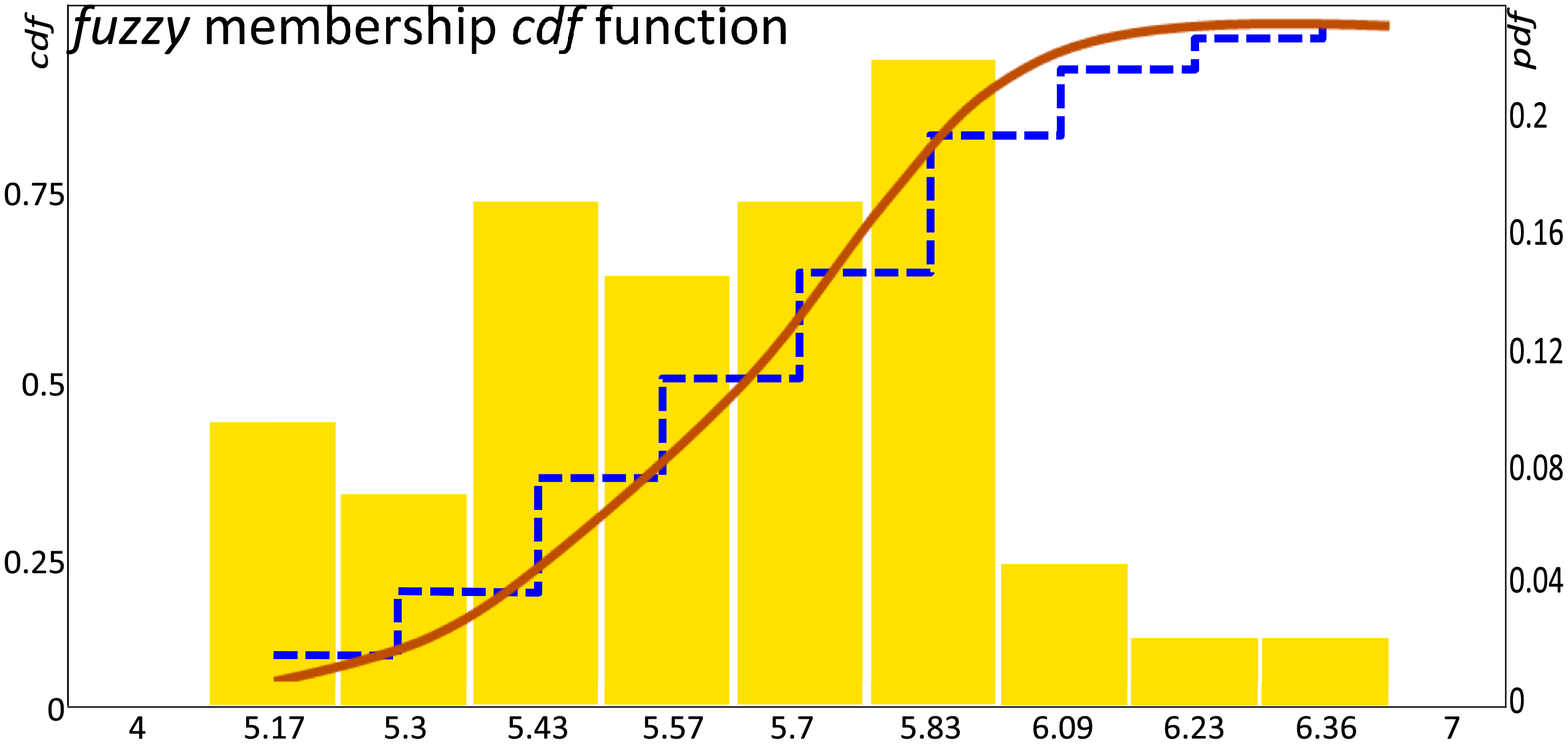}}\\
\subfigure[]{\includegraphics[scale=0.16]{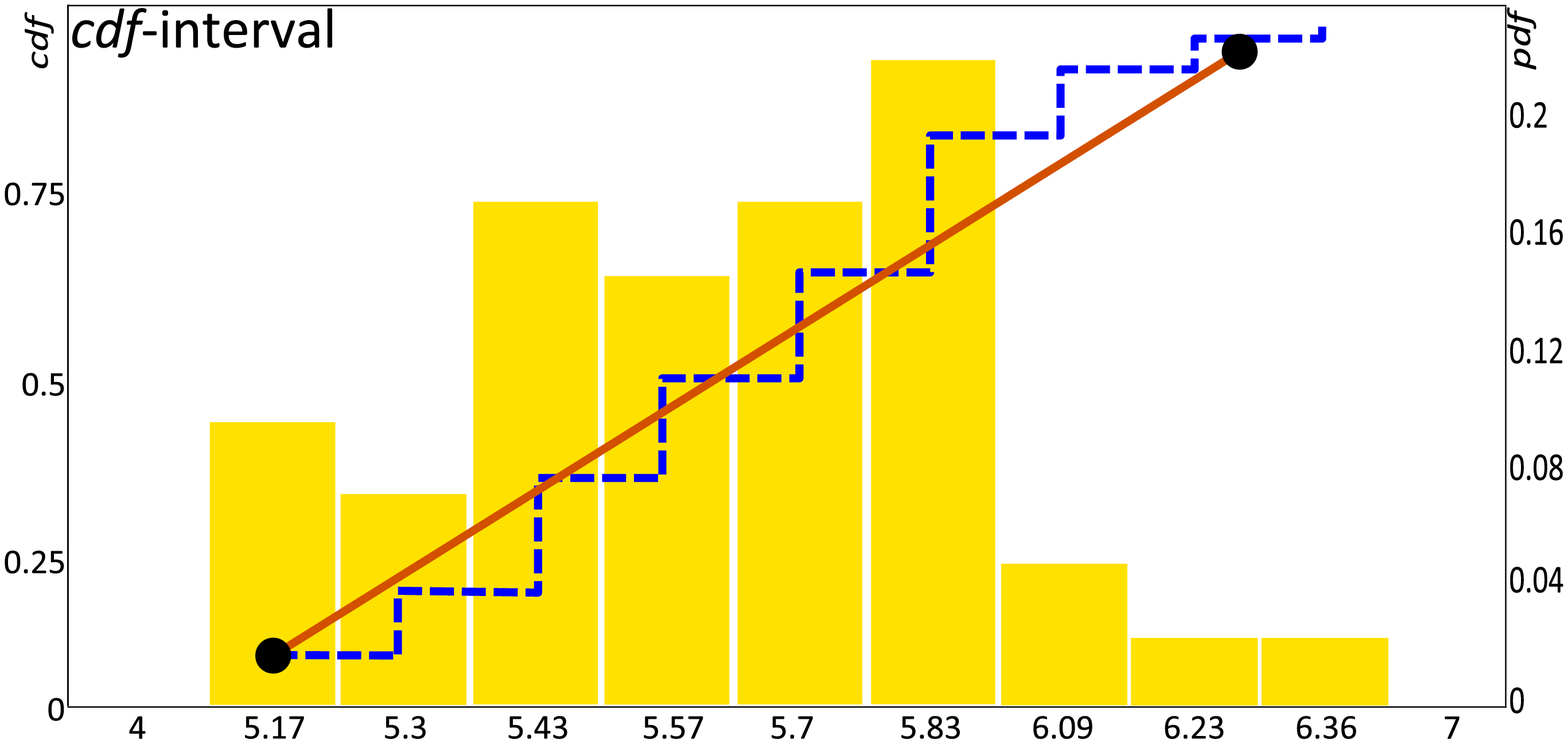}}
\subfigure[]{\includegraphics[scale=0.16]{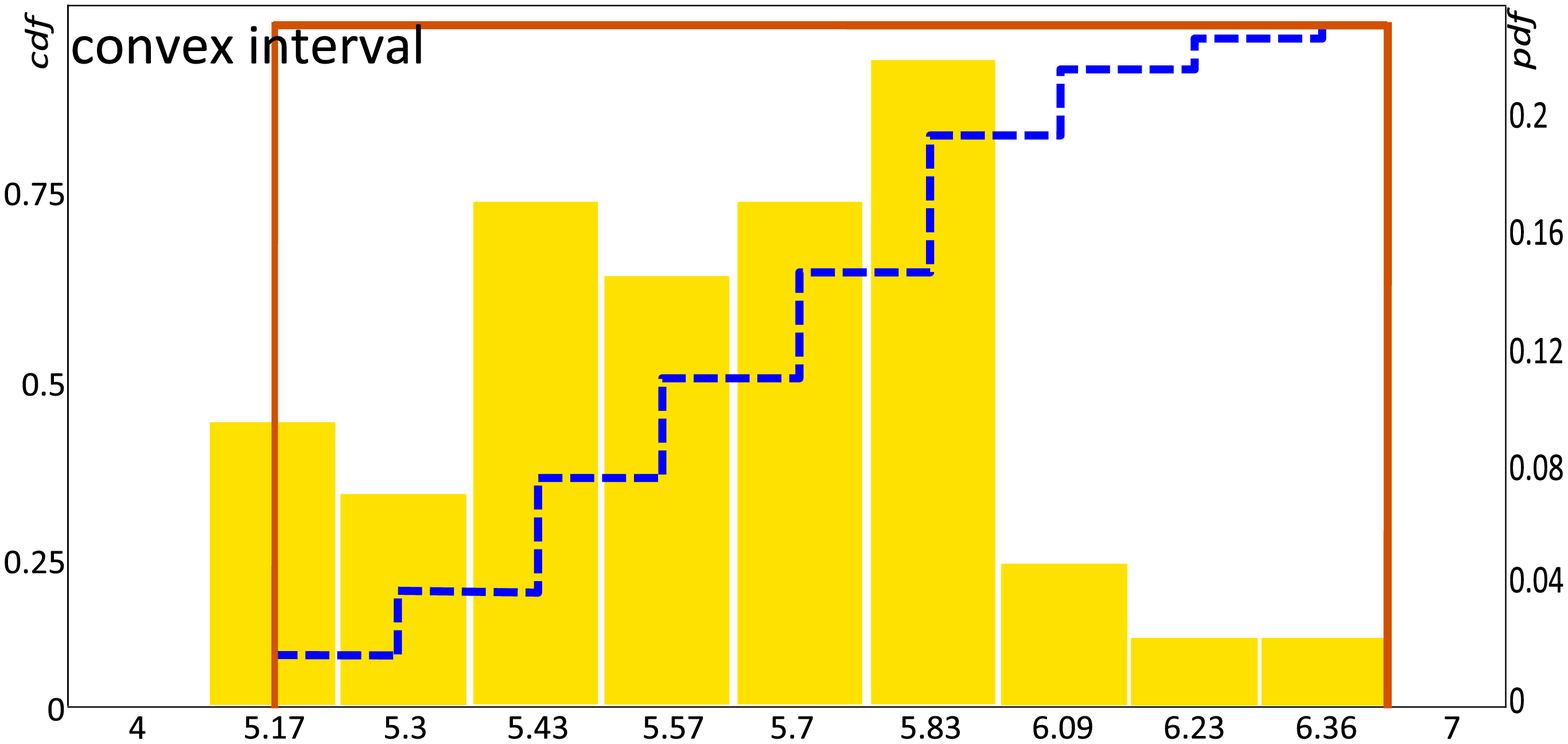}}
\subfigure[]{\includegraphics[scale=0.16]{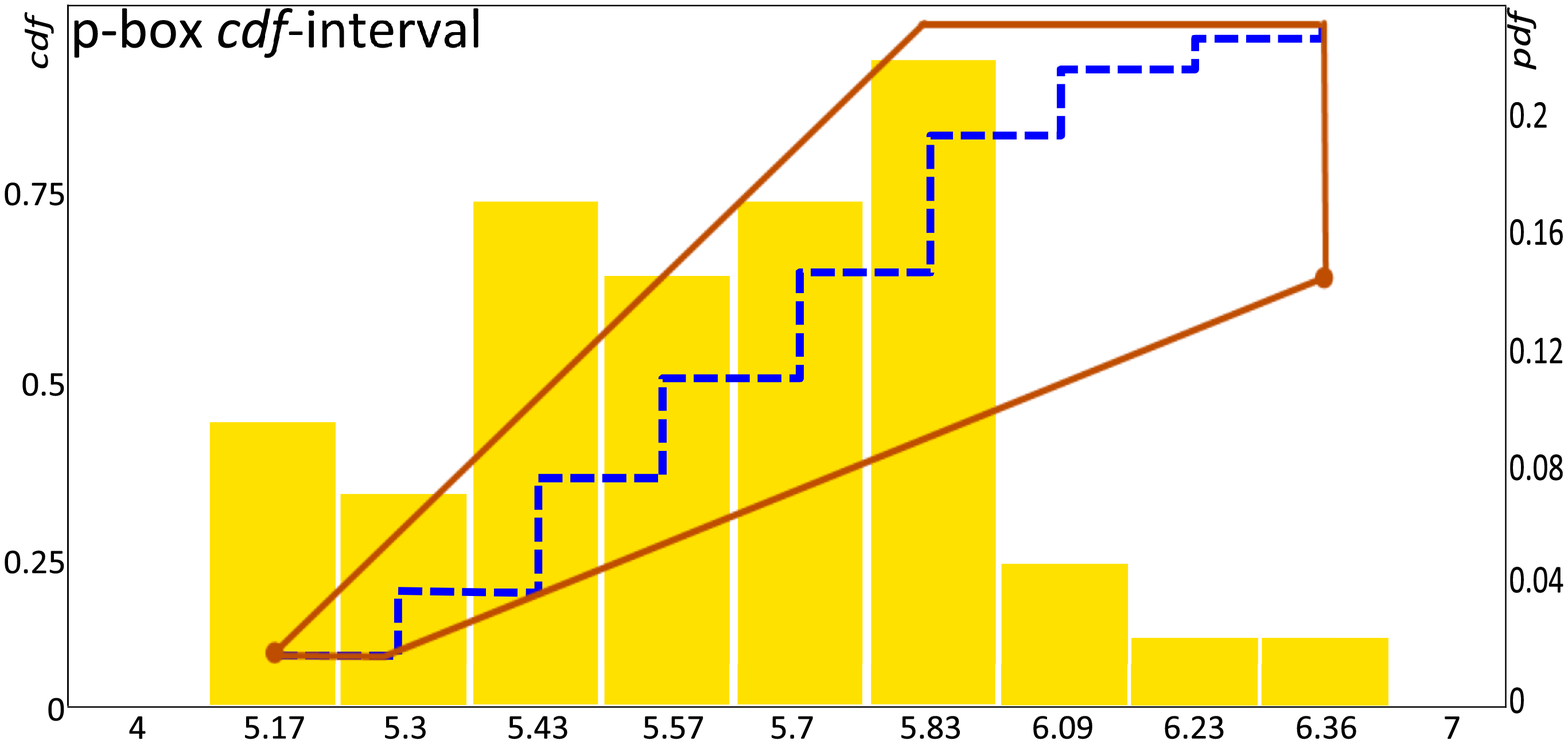}}
\caption{Derived probabilities and {\em cdf} distributions of the steel stud item varying cost} \label{fig:cdfprojection}
\end{figure} 
\paragraph{Projecting the distributions onto the cdf-domain.}
By definition, the {\em cdf} keeps the probabilistic information in an aggregated manner. Information obtained from the measurement process is often discrete and incomplete, hence, when it is projected onto the {\em cdf}-domain, it forms a staircase shape \cite{smith1992approximation}. This is depicted in our running example by the dotted staircase shape in Figure \ref{fig:cdfprojection}. Normal and {\em fuzzy} {\em cdf} distributions are shown by the continuous red curves in Fig. \ref{fig:cdfprojection} (b) and (c). Each is based on an approximation that lacks precise point fitting of the original data whereabouts. Similarly, the {\em cdf}-interval, in Fig. \ref{fig:cdfprojection} (d), approximates the data whereabouts by means of a line connecting the two bounding data values. The convex model representation however shapes a rectangle, illustrated in Fig. \ref{fig:cdfprojection} (e). This rectangle includes all values in the {\em cdf} range $[0,1]$. The convex representation treats data values lying within the interval bounds equally, i.e. it lacks the probabilistic information. The p-box {\em cdf}-interval enforces tighter bounds on the probabilities in the {\em cdf} domain when compared to convex models as depicted in Fig. \ref{fig:cdfprojection} (f). 
\subsection{Constructing the p-box cdf-intervals}\label{subsec:ConstructingpboxIntervals}
Algorithm \ref{alg:pboxdataintervalconstruction} shows the p-box {\em cdf}-interval construction steps. Two parameters are taken into consideration: $\mbox{Arr}[n]$ is an array of $n$ distinct observed and sorted quantile values; whereas the second parameter, $\mbox{\em cdf}[n]$, is the set of their computed {\em cdf} values. The two arrays, together, form the staircase function shape with quantiles stored in $\mbox{Arr}[]$ and {\em cdf} values stored in $\mbox{\em cdf}[]$. Note that a staircase function defines as set of constant values $\mbox{\em cdf}[i]$ over a set of intervals $[\mbox{Arr}[i], \mbox{Arr}[i+1]]$ $\forall i < n$ \cite{smith1992approximation}. Accordingly, the set of upper and lower bounding points forming the staircase function are $\{[\mbox{Arr}[i], \mbox{\em cdf}[i]]\}$ $\forall i, 1 \leq i \leq n$ and $\{[\mbox{Arr}[i+1], \mbox{\em cdf}[i]]\}$ $\forall i, 1 \leq i < n$ respectively. The aim of the algorithm is to envelop those observed points with the highest and lowest possible average probabilistic step increase from the first quantile interval of the staircase function. Issuing the slopes from this specific interval is sufficient to compute the bounds due to the {\em cdf} monotonic property. 
\begin{figure}
\figrule
\begin{center}
\begin{verbatim}
procedure ConstructPBOXCDFIntervalBounds(Arr[n], cdf[n])
// compute the list of slopes between the observed points in the {\em cdf}-domain  
1: j <- 0
2: for i = 2 to n do 
3:    slopeslb[j] <- (cdf[i] - cdf[1]) / (Arr[i] - Arr[1])
4:    slopesub[j] <- (cdf[i-1] - cdf[1]) / (Arr[i] - Arr[2])
// find the most increasing lower bound slope O(nlog(n)) \\
5: Sxl <-� getmax(slopeslb) 
// find the least increasing upper bound slope O(nlog(n)) \\
6: Sxu <-� getmin(slopesub)  
// get the lower bound point \\
7: a <- Arr[1], Fa <-�cdf[1], Sa <- Sxl
// get the upper bound point by projecting the maximum observed quantile \\ 
// onto the upper bound slope \\
8: b <-�Arr[n], Fb <-�Sxu(Arr[n]-Arr[2]) + cdf[1], Sb <- Sxu 
// return the p-box {\em cdf}-interval \\
9: [(a, Fa, Sa),(b, Fb, Sb)] 
\end{verbatim}
\caption{Data interval bounds construction}
\label{alg:pboxdataintervalconstruction}
\end{center}
\figrule
\end{figure}
A {\em cdf} slope, by definition, is the average step value that indicates how the probability distribution increases. Algorithm in Fig. \ref{alg:pboxdataintervalconstruction} starts by computing $2n$ slopes issued from the $2$ points, specified as ($\mbox{Arr}[1]$,$\mbox{\em cdf}[1]$) and ($\mbox{Arr}[2]$,$\mbox{\em cdf}[1]$), and destined to all other points in the {\em cdf}-domain. This is to calculate the list of possible average step values between the observed staircase bounding points. Slopes are then sorted to extract the steepest line and the flatest line. The geometric area under the line, computed by the integral, determines the dominated (dominating) {\em cdf} distribution with maximum (minimum) area as indicated by the stochastic dominance property that is used to order probabilities. Accordingly, the lower bound in the {\em cdf} domain is the fastest increasing line slope and issued from the $1^{st}$ quantile observation, and vise versa the upper bound is the least increasing line slope and issued from the maximum quantile value having the minimum observed {\em cdf} value. This is to guarantee the full encapsulation of all the measured data between the two bounding distributions, each is shaping a line, and together they are ordered by means of the probabilistic stochastic ordering. Algorithm in Fig. \ref{alg:pboxdataintervalconstruction} is correct with time complexity $O(n log(n))$. The proof is omitted for space reason.

The red box in Fig. \ref{fig:cdfprojection} (f) illustrates the p-box {\em cdf}-interval, as opposed to the red line representing the {\em cdf}-interval, constructed for the same set of observations using the `ConstrucIntervalBounds` algorithm proposed in \cite{saad2010constraint}. The {\em cdf}-interval of the same running example is bounded by the points $(5.17,0.1)$ and $(6.25,0.98)$, while the p-box {\em cdf}-interval representation is bounded by the points $(5.17,0.1)$ and $(6.36,0.7)$, each lying on a bounding {\em cdf} uniform distribution with slopes $1.2$ and $0.57$ respectively.
\subsection{Interpretation of the confidence interval ${\bf I}$}
We formally describe the p-box {\em cdf}-interval structure which is bounding the observed data as shown in Algorithm \ref{alg:pboxdataintervalconstruction}. The theoretical algebraic representation of an interval of points is specified by ${\bf I}=[p_a,p_b]$, where $p_a$ and $p_b$ are the extreme points which bound the p-box {\em cdf}-interval. Throughout this paper, we assume that data takes its value in the set of real numbers $\mathbb{R}$, denoted by $a$,$b$,$c$. Data points are denoted by $p$,$q$,$r$, possibly subscripted by a data value (quantile). 
\paragraph{The p-box cdf-interval ${\bf I}=$[$p_a,q_b$].}
One can see that this interval approach does not aim at approximating
the curve but rather enclosing it in a reliable manner. The complete envelopment is exerted by means of the uniform {\em cdf}-bounds, which are depicted by the red curves in Fig. \ref{fig:cdfprojection} (f). It is impossible to find a point that exists outside the formed interval bounds. The {\em cdf} bounds are chosen to have a uniform distribution due to its linear computational complexity. Each is represented by a line with a slope ($S^p_a$,$S^q_b$) issued from one of the extreme quantiles ($a$,$b$). Storing the full information of each bound is sufficient to restore the designated interval assignment. Bounds are denoted by triplet points, in the $2$D space, to guarantee the full information on: the extreme quantile values observed ($a$,$b$); the {\em cdf}-value of each quantile projected onto its corresponding bounding distribution ($F^p_a$,$F^q_b$); and the degree of steepness formed by the uniform distributions ($S^p_a$,$S^q_b$). The uniform {\em cdf}-distribution has a line shape with a slope indicating how the probabilistic values accumulate for successive quantiles. Accordingly, the p-box {\em cdf}-interval bounding points representation: $p_a = (a,F^p_a,S^p_a)$ and $q_b = (b,F^q_b,S^q_b)$. The p-box {\em cdf}-intervals triplet points are ordered in $\mathcal{U}$, where $\mathcal{U}$ is a partial order set defined over $\mathbb{R}\times [0,1] \times \mathbb{R}^+$ with an ordering operator $\preccurlyeq_\mathcal{U}$. 
\begin{definition}
$S^p_x$ is the slope of a given cdf-distribution; it signifies the average step probabilistic value. For a given uniform cdf-distribution 
\begin{equation}
S^p_x = \frac{{F_b - F_a}}{b - a}, \forall a \leq x \leq b
\end{equation} 
The average step value, denoted as $S^p_x$, derives the probabilistic values of consequent quantiles on the real domain. 
\end{definition}
Plotting a point $p_x$ within the p-box {\em cdf}-interval deduces bounds on its possible chances of occurrence.
\begin{definition}\label{def:FPBOXpx} $F_x^I$ is the interval of cdf values obtained when $p_x$ is projected onto the {\em p-box} cdf bounds. For a point $p_x \in {\bf I}$ denoted as $p_x=(x, F_x^p, S_x^p)$ and $p_a \preccurlyeq_\mathcal{U} p_x \preccurlyeq_\mathcal{U} q_b$ 
\begin{equation}\label{eq:pointordering}
a<x<b,~\mbox{and}~ F_b^{q`} \geq F_x^I \geq F_a^{p`} ~\mbox{and}~ S^p_a \geq S^p_x \geq S^q_b
\end{equation} 
\end{definition}
$F^{p`}_a$ and $F^{q`}_b$ are the possible maximum and minimum {\em cdf} values $p_x$ can take; both are computed by projecting the point $p_x$ onto the {\em cdf} distributions passing through real points $a$ and $b$ respectively. They are derived using the following linear projections, computed in $O(1)$ complexity:
\begin{equation}\label{eq:boundingcdfvalues}
F_a^{p`} = min(S_a^p(x-a) + F_a^p,1) ~~~~ \mbox{and} ~~~~ F_b^{p`} = max(F_b^p - S_b^p(b-x),0) \nonumber
\end{equation}
Equation \ref{eq:boundingcdfvalues} guarantees the probabilistic feature of the {\em cdf}-function by restricting its aggregated value from exceeding the value $1$ and having negative values below $0$. 
\begin{example}
${\bf I} = [(5.17, 0.1, 1.2), (6.36, 0.7, 0.57)]$ is the p-box {\em cdf}-interval of the cost/item in Example \ref{ex:costperitemdataobservation}. Suppose that $x_i=5.5$, its {\em cdf}-bound values $F_x^I = [0.2,0.5]$. This means that the possible chance of the value to be at most $5.5$ is between $20\%$ and $50\%$, with an average step probabilistic value between $0.57$ and $1.2$. Note that this interval is opposed to only one approximated value $F_x = 0.37$ in the {\em cdf}-intervals representation proposed in \cite{saad2010constraint}, the {\em fuzzy} {\em cdf} value $F_x = 0.31$ and its Normal {\em cdf} value is $F_x = 0.42$. Note that convex models do not enforce any probabilistic bounds, accordingly, $x_i=5.5$ has a {\em cdf} $F^I_x \in [0,1]$.  
\end{example}
\section{Constraint reasoning}\label{sec:constraintreasoning}
In the CP paradigm, relations between variables are specified as constraints. A set of rules and algebraic semantics, defined over the list of constraints, formalize the reasoning about the problem. As a fundamental language component in the Constraint Logic Programming (CLP), these set of rules, with a syntax of definite clauses, form the language scheme \cite{jaffar1987}. The constraint solving scheme is intuitively and efficiently utilized in the reasoning over the computation domain. The scheme formally attempts at assigning to variables a suitable domain of discourse equipped with an equality theory together with a least and a greatest model of fix-point semantics. Starting from an initial state the reasoning scheme follows a local consistency technique which attempts at constraining each variable over the p-box {\em cdf}-interval domain while excluding values which do not belong to the feasible solution. An implementation of the constraint system was established as a separate module in the ECL{\em $^i$}PS{\em $^e$} constraint programming environment \cite{eclipse}. ECL{\em $^i$}PS{\em $^e$}  provides two major components to build the solver: an attributed variable data structure and a suspension handling mechanism. Fundamentally, attributed variables are specific data structures which attach more than one data type. Together they permit for a new definition of unification which extends the well-known Prolog unification \cite{lehuitouze1990,holzbaur92}.  A p-box {\em cdf}-interval point is implemented in an attributed variable data structure with three main components: quantile, {\em cdf} value and slope. Whilst constraints suspension handling is a highly flexible mechanism that aims at controlling user defined atomic goals. This is achieved by waiting for user-defined conditions to trigger specific goals.
 
\begin{figure}[hc]
\centering
\subfigure[]{\includegraphics[scale=0.2]{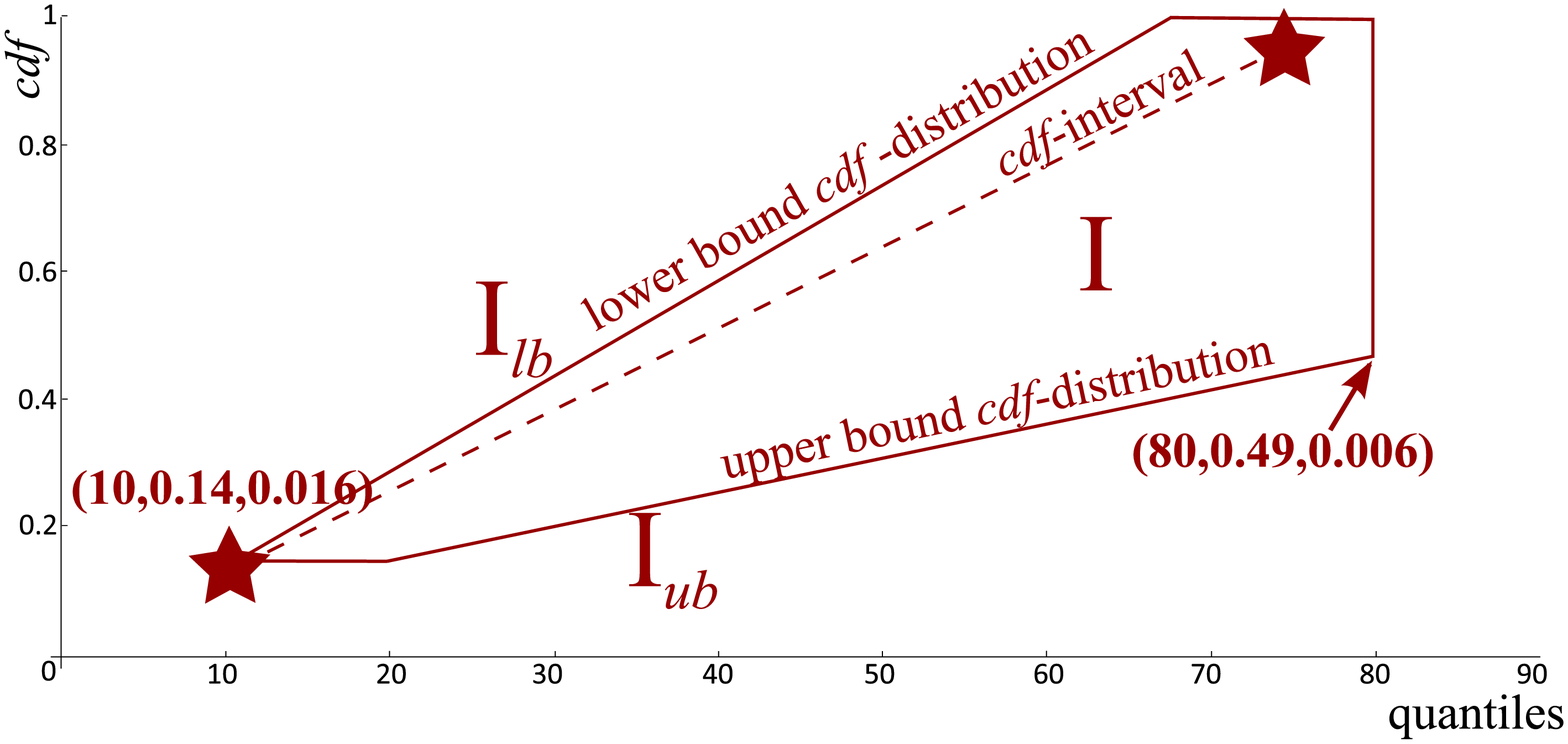}}
\subfigure[]{\includegraphics[scale=0.2]{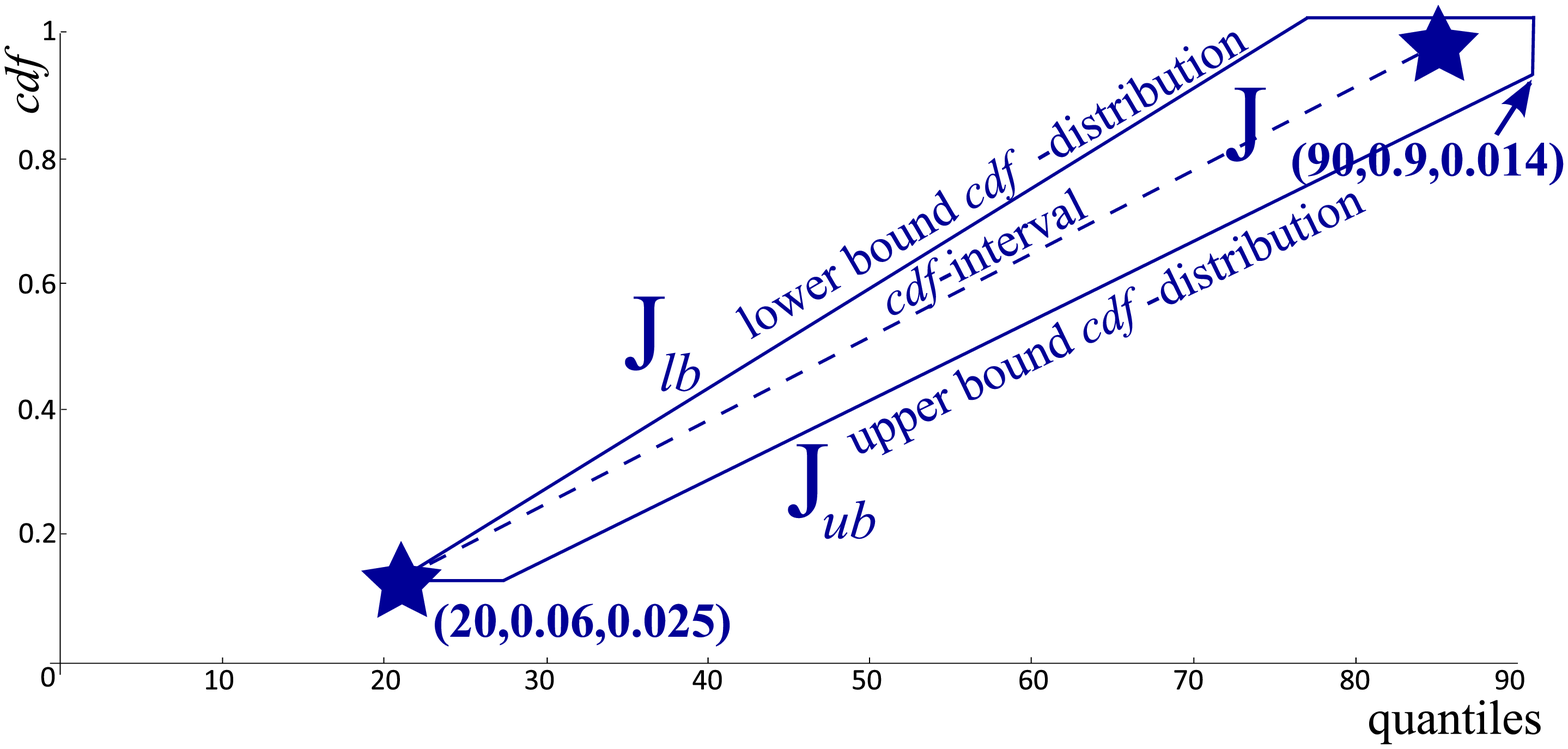}}\\
\subfigure[]{\includegraphics[scale=0.2]{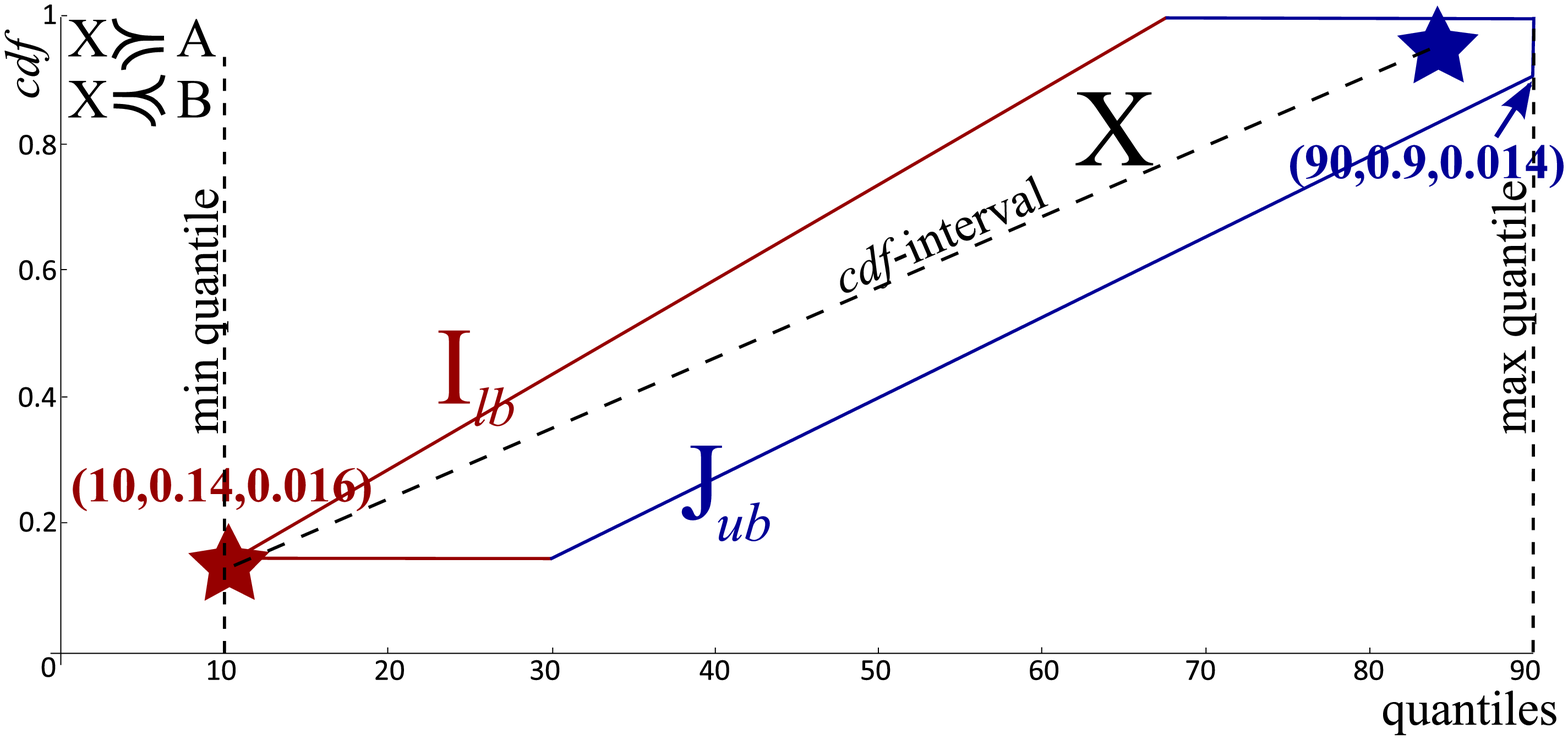}}
\subfigure[]{\includegraphics[scale=0.2]{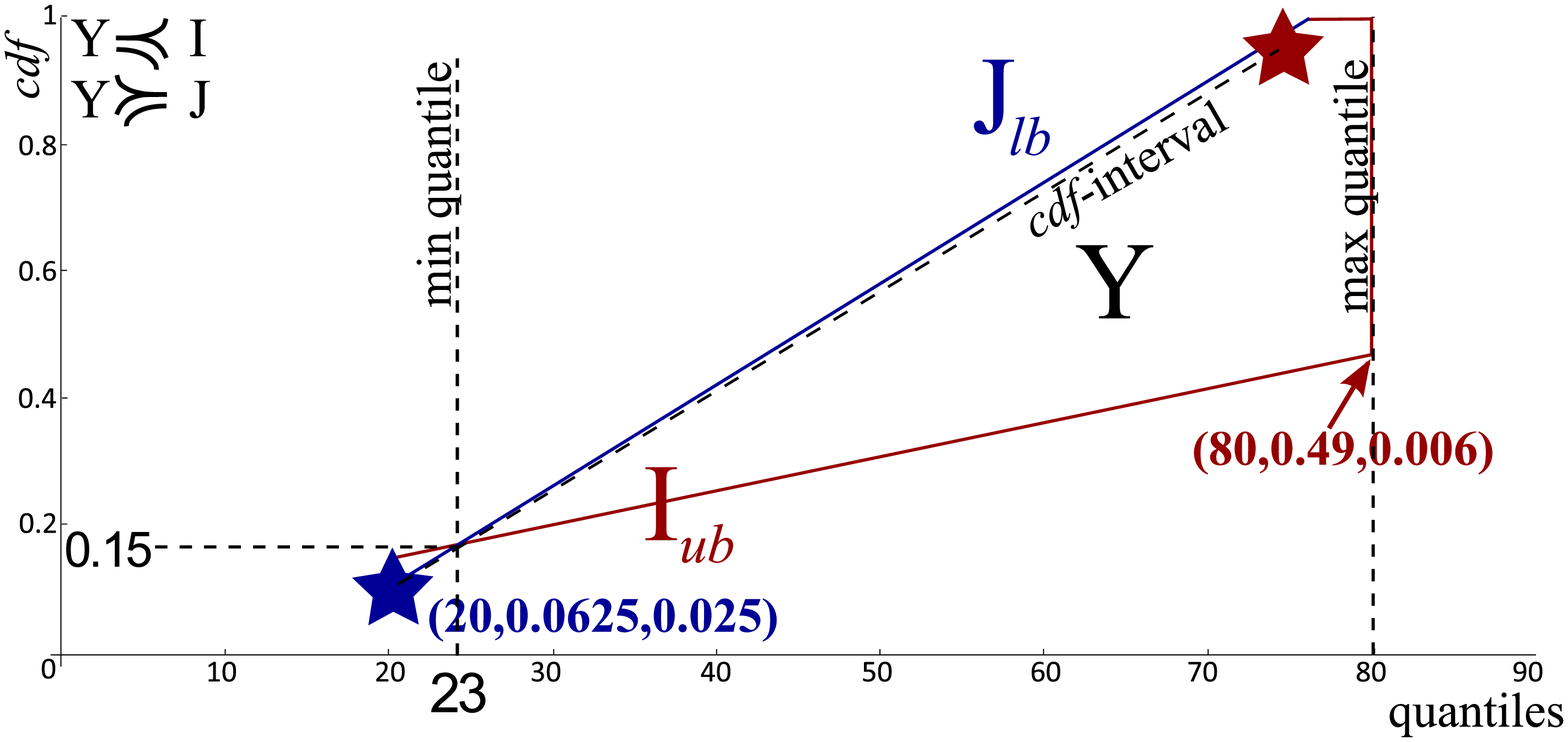}}
\caption{Ordering constraint execution} \label{fig:orderingconstraint}
\end{figure}

Implemented rules in our solver infer the local consistency in the p-box {\em cdf}-interval domains of the binary equality and ordering constraints $\{ =, \preccurlyeq_\mathcal{U} \}$, and that of the ternary arithmetic constraints $\{ +_\mathcal{U}, -_\mathcal{U}, \times_\mathcal{U}, \div_\mathcal{U} \}$. Operations, in the solver, are exerted first as real interval computations, and then they are projected onto the {\em cdf} domain using a linear computation, as shown in Definition \ref{def:FPBOXpx}. In this section we demonstrate how the ordering and the ternary addition constraints infer the local consistency over the variable domains of $X$, $Y$, and $Z$ assuming that their initial bindings are $I = [p_a,p_b]$, $J = [q_c,q_d]$ and $K = [r_e,r_f]$ respectively. The ternary multiplication, subtraction and division constraints are implemented in the same way.
\paragraph{Ordering constraint $X \preccurlyeq_{\mathcal{U}} Y$}. To infer the local consistency of the binary ordering constraint, we extend the lower {\em cdf}-bound of $X$ and contract the upper {\em cdf}-bound of $Y$. The ordering constraint is defined by the following rule: 
\begin{eqnarray}
\frac{p_b\textquotesingle = glb(p_b,q_d),{q_c}\textquotesingle = lub(p_a,q_c)}{\{X \in {\bf I}, Y \in {\bf J},X \preccurlyeq_\mathcal{U} Y\} \longmapsto \{ X \in [p_a,p_b\textquotesingle ], Y \in [q_c\textquotesingle ,q_d], X \preccurlyeq_\mathcal{U} Y\}} \nonumber
\end{eqnarray} 
To achieve the local consistency, the ordering constraint $\preccurlyeq_\mathcal{U}$ updates the upper bound of the variable $X$ domain to $glb(p_b,q_d)$, which is the greatest lower bound of the two points, i.e. the point preceeding the two on the partially ordered set lattice $\mathcal{U}$. And vice versa, the lower bound of $Y$ is updated to $lub(p_a,q_c)$ (the least upper bound of the two points). 
\begin{example}
Let ${\bf I}$ and ${\bf J}$ be two p-box {\em cdf}-interval domains. ${\bf I} = [(10,0.14,0.016),(80,0.49,0.06)]$ and ${\bf J} = [(20,0.06,0.025),(90,0.9,0.014)]$. The effect of applying the set of constraints $X \succcurlyeq_{\mathcal{U}} {\bf I}$ and $X \preccurlyeq_{\mathcal{U}} {\bf J}$, prunes the domain of $X$. As a result, the variable $X$ is bounded by the lower bound of ${\bf I}$ and by the upper bound of ${\bf J}$: $X \in [(10,0.14,0.016),(90,0.9,0.014)]$ as shown in Fig. \ref{fig:orderingconstraint} (a). Clearly the obtained domain of $X$, in this example, preserves the convex property of the p-box {\em cdf}-intervals. 
Let $Y$ be subject to the domain pruning using the set of constraints: $Y \preccurlyeq_{\mathcal{U}} {\bf I}$ and $Y \succcurlyeq_{\mathcal{U}} {\bf J}$. As a result, $Y$ should be bounded by the lower bound of $\bf{J}$ and the upper bound of $\bf{I}$. However, in this case, at lower quantiles $\leq 23$, the upper bound distribution of $\bf{I}$ preceeds the lower bound of $\bf{J}$. The fact that conflicts the stochastic dominance property of a p-box {\em cdf}-interval domain. In order to resolve this conflict, the real bounds of $Y$ are further pruned to the point of the probability intersection $=23$. 
\end{example}
\vspace{-1\baselineskip}
\paragraph{Ternary addition constraints $X +_\mathcal{U} Y = Z$}. The addition operation is implemented by summing up pair of points, defined in the $2$D space and located within the p-box {\em cdf}-interval bounds which enclose the domain ranges of $X$ and $Y$. This addition operation is linear. It is convex and can be computed from the end points of the domains involved in the addition. The p-box {\em cdf}-domain of $Z$ is updated to envelop all points defined in that range. To infer about the {\em cdf} ternary addition constraint we use the following rule:  
\begin{eqnarray}\label{eq:addconstraint11}
\frac{r_f\textquotesingle = ( ub_+,F^{I+J}_{ub_+},S^{I+J}_{ub_+} ),r_e\textquotesingle = ( lb_+,F^{I+J}_{lb_+},S^{I+J}_{lb_+})}{\{X \in {\bf I}, Y \in {\bf J}, Z \in {\bf K}, Z = X +_\mathcal{U} Y\} \longmapsto \{ X \in {\bf I}, Y \in {\bf J}, Z \in [r_e\textquotesingle ,r_f\textquotesingle], Z = X +_\mathcal{U} Y\}}
\end{eqnarray} \\
\begin{eqnarray}\label{eq:addconstraint12}
\frac{p_b\textquotesingle = ( ub_-,F^{K-J}_{ub_-},S^{K-J}_{ub_-} ),p_a\textquotesingle = ( lb_-,F^{K-J}_{lb_-},S^{K-J}_{lb_-})}{\{X \in {\bf I}, Y \in {\bf J}, Z \in {\bf K}, X = Z -_\mathcal{U} Y\} \longmapsto \{ X \in [p_a\textquotesingle ,p_b\textquotesingle], Y \in {\bf J}, Z \in {\bf K}, Z = Z -_\mathcal{U} Y\}} 
\end{eqnarray} 
The projection onto the $Y$ domain is symmetrical. The p-box {\em cdf} ternary addition inference rule is exerted on the variable domains involved in the relation $Z = X +_\mathcal{U} Y$. The domain of $Z$ is updated with the addition of the two interval domains $I$ and $J$ which yields a lower bound $(lb_+,F^{I+J}_{lb_+},S^{I+J}_{lb_+})$ and an upper bound $(ub_+,F^{I+J}_{ub_+},S^{I+J}_{ub_+})$. $lb_+$ and $ub_+$ are the bounds of the arithmetic addition exerted on the real domain $\mathbb{R}$. $(F^{I+J}_{lb_+},S^{I+J}_{lb_+})$ and $(F^{I+J}_{ub_+},S^{I+J}_{ub_+})$ are the bounding {\em cdf} distributions, each is obtained by means of a linear equation that is proposed in \cite{saadcdf}, and which is derived using the approach in \cite{glen2004computing}. The domain of $Z$ is pruned by intersection the new bounding points $[r_e\textquotesingle,r_f\textquotesingle]$, resulting from the p-box {\em cdf}-intervals addition operation, with the initial binding of $Z$. Since three variables are involves in the ternary addition, domains of $X$ and $Y$ are pruned using rule \ref{eq:addconstraint12}. The p-box {\em cdf}-interval subtraction is exerted linearly over the bounding points of $K - J$ and $K - I$. $(lb_-,F^{K-J}_{lb_-},S^{K-J}_{lb_-})$ and $(ub_-,F^{K-J}_{ub_-},S^{K-J}_{ub_-})$ are the resuting bounds defined over $\mathcal{U}$ and they are intersected with the initial binding of $X$. Similarly the domain of $Y$ is pruned. This operation is exerted multiple times until the constraint is stabilized, i.e. no further pruning is taking place and the system of constraint is preserving its local consistency.

The ternary addition constraint exerted on p-box {\em cdf}-interval domains is a simple addition computation since it adopts the real-interval arithmetics which are then projected linearly onto the {\em cdf} domain. This operation is opposed to the {\em fuzzy} extended addition operation adopted in the constraint reasoning utilized in the possibilistic domain \cite{dutta2005single,petrovic1996fuzzy}, and to the Normal probabilistic addition which has a high computation complexity that is due to the Normal distribution shape \cite{glen2004computing}. 
\section{Empirical evaluation}\label{sec:empiricalevaluation}
We use, as a case study, an inventory management problem. We adopt in our evaluation the model proposed by \cite{tarim2004}. The key idea is to schedule
ahead replenishement periods and find the optimal order sizes which achieve a minimum total manufacturing cost. A reorder point $\delta_t$ with order size $X_t$ should meet customer demands $d_t$ up to the next point of replenishment with an adequate inventory level $I_t$. 
\begin{definition}\label{def:pboxcdfimodel}
An inventory management model defined over a time horizon of $N$ cycles is 
\begin{eqnarray}\label{eq:pboxcdfimodel}
\mbox{minimize   }~~~~TC = \sum^{N}_{t=1}(a\delta_t + hI_t + vX_t) \nonumber \\
\mbox{subject to    }~~~~\delta_t = \left\{ \begin{array}{cc}
1 & \mbox{if}~~X_t >0 \\
0 & \mbox{otherwise} \\
\end{array} \right\} \nonumber \\
I_t = I_0 + \Sigma^{t}_{i=1} ( X_i - d_i ) \nonumber \\
X_t, I_t \geqslant 0,~~ t = 0,1,...,N  
\end{eqnarray}
\end{definition}
The problem is an optimization problem that seeks the minimization of the total cost $TC$ which constitutes of three components: the cost of replenishment which is defined by the ordering cost $a$ multiplied by the number of times a replenishment takes place $\sum^{N}_{t=1}\delta_t$; the holding cost which  depends on the depreciation cost $h$ and the level of the inventory observed in a given cycle $I_t$; and the purchase cost which is the reorder quantity $X_t$ multiplied by the varying cost/item $v$. The model is studied over a time horizon of $N$ cycles. $\delta_t$ is $1$, when an order is issued and $0$ otherwise. The inventory level $I_t$ for a given cycle is the difference between the ordered items $X_t$ and those which are consumed $d_t$. $I_0$ is the initial inventory level. From this model, one can observe that all cost components depend totally on fluctuating and unpredictable variables especially in the real-life version of the problem. This is due to the unpredictability of customer demands and the variability of the cost/item. Accordingly, this model perfectly fits our evaluation criteria: comparing the behavior of the models when the environment is uncertain.    
\paragraph{Information realized in the solution set.}
We test the model for a randomly distributed monthly demands. Table \ref{tab:solcomparison} shows the average demand per cycle for a time horizon $N = 10$ cycles. We build a p-box {\em cdf}-interval for each average demand value since it is given from a list of customer demand observations over the years. The construction of the p-box {\em cdf}-interval representation follows Algorithm \ref{alg:pboxdataintervalconstruction}. Clearly, {\em fuzzy} and probabilistic models are based on the listed average values. The two models set assumptions on the shape of the probability distribution adopted, as pointed out in Section \ref{sec:datarepresentation}. We then develop the intervals of the cost components. Example \ref{ex:costperitemdataobservation} demonstrates how to deduce the input varying cost/item observed for $12$ months.
We implement the model defined in Equation \ref{eq:pboxcdfimodel}. The input customer demands and cost components are represented as p-box {\em cdf}-intervals. We start the problem with an empty initial inventory. The set of addition and equality constraints are employed in the p-box {\em cdf}-interval domain. Constraints are triggered until stabilized and consistency is reached by means of the inference rules defined in Section \ref{sec:constraintreasoning}. The solver suggests $2$ to $5$ replenishment periods, with a total holding cost $[(8.5, 0.83, 4.4E-04),(137.98, 0.039, 7.5e-5)]$ and a total manufacturing cost $[(2739.6, 0.8, 3.3E-04), (6483.2, 0.03, 6.2e-5)]$. This output is opposed to $6$ replenishment periods realized by the {\em fuzzy} and the probabilistic models, as shown in Table \ref{tab:solcomparison}, with a total holding cost $\$53.5$ and $\$52.05$ and a total manufacturing cost $\$3868.5$ and $\$3828.93$ respectively.
\begin{figure}[hc]
\centering
\subfigure{\includegraphics[scale=0.25]{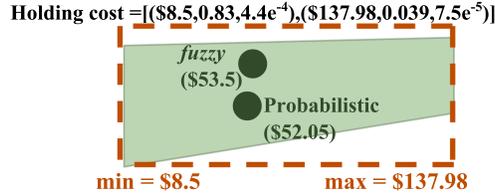}}
\caption{Output solutions for holding cost} \label{fig:outputsolution}
\end{figure}\vspace{-1\baselineskip}
We have successfully added more value to the solution set obtained due to the propagation techniques applied in the p-box {\em cdf}-intervals domain. Fig. \ref{fig:outputsolution} illustrates a comparison between the output holding cost obtained from the models under consideration. The p-box {\em cdf}-interval graphical representation of the cost is depicted by the shaded region and their bounds in the convex models are illustrated by the dotted rectangles. Clearly, the solution set obtained from the p-box {\em cdf}-intervals model, when compared with the outcome of the convex model, realized an additional knowledge (i.e. tighter bounds in the {\em cdf} domain). This solution set is opposed to a one value proposed as $\$53.5$ by the {\em fuzzy} and as $\$52.05$ by the probabilistic models. Output solution point suggested by the latter models can, sometime, mislead or deviate the decision making. This is because their distributions are built, from the begining, on approximating the actual observed distribution.
\begin{center}
\begin{table}
\centering
\scriptsize 
\begin{tabular}{lrrrrrrrrrr}
\hline 
Average $d_t$  & $26$ & $36$ & $23$ & $28$ & $32$ & $30$ & $29$ & $37$ & $25$ & $34$\\
Lower bound $d_t$  & $25.6$ & $34.7$ & $22.5$ & $27.1$ & $31.7$ & $29.6$ & $28.6$ & $36.2$ & $24$ & $33.2$\\
Upper bound $d_t$  & $26.9$ & $36.8$ & $23.9$ & $28.4$ & $33$ & $31.5$ & $29.9$ & $37.9$ & $25.4$ & $34.5$\\
\cmidrule{2-11}
Probabilistic $\delta_t$  & $1$ & $1$ & $1$ & $0$ & $1$ & $0$ & $1$ & $0$ & $1$ & $0$\\
PBOX $\delta_t$  & $[1, 1]$ & $[0, 0]$ & $[0, 0]$ & $[0, 1]$ & $[0, 0]$ & $[0, 1]$ & $[0, 1]$ & $[0, 0]$ & $[0, 0]$ & $[0, 1]$\\
\hline 
\end{tabular}
\caption{$d_t$ and $\delta_t$ over a time horizon of $10$ cycles}
\label{tab:solcomparison}
\end{table} \vspace{-1\baselineskip}
\end{center}
\paragraph{Model tractability.}
We adopt the data corpus introduced by \cite{tarim2006}. They generated $4$ types of randomly distributed demand data sets. Customer demands are varied over the time horizon ($t$ is the cycle number) using the following equations:
\begin{itemize}
\item P1 set (general trend): demand distribution mean value per cycle is \\$50(1+ \sin(\pi t/6))$ 
\item P2 set (positive trend): demand distribution mean value per cycle is \\$50(1+ \sin(\pi t/6))+t$
\item P3 set (negative trend): demand distribution mean value per cycle is \\$50(1+ \sin(\pi t/6))+ (52 - t)$
\item P4 set (life-cycle trend): demand distribution mean value per cycle is \\$50(1+ \sin(\pi t/6)) + \min (t,52-t)$
\end{itemize}
We run the different models for high values of $t$ ($t \geq 30$).
\begin{table}
\begin{minipage}{\textwidth}
\scriptsize 
\begin{tabular}{lrrrrrrrrr}
\hline\hline 
time horizon & $t=30$ & $t=32$ & $t=34$ & $t=36$ & $t=38$ & $t=40$ & $t=42$ & $t=44$ & $t=46$\\
\hline 
&  \multicolumn{9}{c}{\bf{P1 set}} \\
\hline 
Stochastic & $4599.65$ & $5442.04$ & $6355.23$ & $ $ & $ $ & $ $ & $ $ & $ $ & $ $ \\
Probabilistic & $1882.5$ & $1710.91$ & $2207.96$ & $6557.76$ & $ $ & $ $ & $ $ & $ $ & $ $ \\
Fuzzy  & $1138.5$ & $1228.8$ & $1479.68$ & $1697.76$ & $1869.98$ & $2129.6$ & $2328.48$ & $5265.93$ & $ $ \\
CDF & $1244.6$ & $865.78$ & $642.75$ & $891.5$ & $1130$ & $1351.67$ & $2289.59$ & $2340.78$ & $ $ \\
PBOX  & $675.77$ & $586.81$ & $874.12$ & $1110.59$ & $1256.86$ & $1955.72$ & $2119.47$ & $ $ & $ $ \\
Convex & $1111.26$ & $432.88$ & $553.48$ & $778.28$ & $961.24$ & $1088.4$ & $1800.23$ & $1844.06$ & $1828$ \\
\hline 
&  \multicolumn{9}{c}{\bf{P2 set}} \\
\hline 
Stochastic & $4650.8$ & $5502.57$ & $6425.92$ & $ $ & $ $ & $ $ & $ $ & $ $ & $ $ \\
Probabilistic & $1422$ & $3242.4$ & $5248.25$ & $ $ & $ $ & $ $ & $ $ & $ $ & $ $ \\
Fuzzy & $1620$ & $2088.96$ & $2653.02$ & $3311.28$ & $3869.92$ & $5136$ & $6615$ & $ $ & $ $ \\
CDF & $1465.45$ & $775.08$ & $538.88$ & $854.55$ & $1285.74$ & $1922.06$ & $2102.92$ & $ $ & $ $ \\
PBOX & $1376.66$ & $669.89$ & $520.13$ & $813.36$ & $1211.82$ & $1663.99$ & $1985.7$ & $ $ & $ $ \\
Convex & $1238.79$ & $440.33$ & $468.82$ & $693.04$ & $1095.12$ & $1371.14$ & $1814.8$ & $ $ & $ $ \\
\hline 
&  \multicolumn{9}{c}{\bf{P3 set}} \\
\hline 
Stochastic & $4590.34$ & $5431.04$ & $6342.38$ & $ $ & $ $ & $ $ & $ $ & $ $ & $ $ \\
Probabilistic & $1773.75$ & $2444.8$ & $4722.27$ & $6156$ & $ $ & $ $ & $ $ & $ $ & $ $ \\
Fuzzy & $1696.5$ & $2216.96$ & $3034.5$ & $3777.85$ & $4194.83$ & $5192$ & $7003.09$ & $ $ & $ $ \\
CDF & $1195.14$ & $888.15$ & $622.29$ & $1073.09$ & $1372.47$ & $1775.58$ & $2435.39$ & $ $ & $ $ \\
PBOX & $1047.68$ & $840.45$ & $532.45$ & $920$ & $1172.04$ & $1567.14$ & $2147.39$ & $ $ & $ $ \\
Convex & $897.83$ & $743.92$ & $529.05$ & $848.64$ & $1144.34$ & $1548.07$ & $2091.32$ & $ $ & $ $ \\
\hline 
&  \multicolumn{9}{c}{\bf{P4 set}} \\
\hline 
Stochastic & $4604.29$ & $5447.54$ & $6361.65$ & $ $ & $ $ & $ $ & $ $ & $ $ & $ $ \\
Probabilistic & $2259$ & $2672.64$ & $4404.36$ & $ $ & $ $ & $ $ & $ $ & $ $ & $ $ \\
Fuzzy & $1831.5$ & $2319.36$ & $3063.4$ & $3531.6$ & $4534.16$ & $4534.16$ & $6368.04$ & $ $ & $ $ \\
CDF & $1357.18$ & $800.11$ & $605.21$ & $922.54$ & $1127.69$ & $1379.99$ & $1990.82$ & $2051.26$ & $ $ \\
PBOX & $1156.04$ & $664.42$ & $601.99$ & $813.17$ & $1010.49$ & $1186.99$ & $1698.63$ & $1684.34$ & $ $ \\
Convex & $1155.23$ & $442.47$ & $519.8$ & $697.55$ & $968.99$ & $1177.76$ & $1570.67$ & $1449.6$ & $1669$ \\
\hline\hline 
\end{tabular}
\caption{Real-time taken to solve instances for the demand sets: P1, P2, P3 \& P4}
\label{tab:pset}
\vspace{-2\baselineskip}
\end{minipage}
\end{table} 
Table \ref{tab:pset} shows the time taken in seconds by each model to reach a solution for the varying demands in a given time horizon. Timeout is set to $2$ hours. Empty cells in the table demonstrate the failure of the model to solve the problem within the $2$ hours interval. As shown in rows (3,10,17 and 24), stochastic models time-out after a time horizon $t=34$. Clearly they have the most expensive computations because they work on the probability distribution in a pointwise manner. Observing each column in Table \ref{tab:pset}, one can notice the speed of each model to reach a solution for the given problem. Evidently, convex models outperfrom the rest of the models in terms of speed; p-box {\em cdf}-intervals have a closer speed, followed by {\em fuzzy} models, then the probabilistic models. In summary, the p-box {\em cdf}-intervals speed performance is closer to that of the convex models. This means that, the new framework, with minimal overhead, adds up a quantifiable information by imposing tighter bounds on the probability distribution, in a safe and a tractable manner. We claim that applied computations are tractable because they are exerted on the interval bounds, using interval computations, then results are further projected, linearly, onto the {\em cdf} domain. Last but not least, empirical evaluations which we used to test the scalability of the framework support our argument.\vspace{-1\baselineskip} 
\section{Conclusion and future research direction}
In this paper, we propose a novel constraint domain to reason about data with uncertainty. The key idea is to extend convex models with the notion of p-boxes in order to realize aditional quantifiable information on the data whereabouts. To the best of our knowledge, p-boxes have never been implemented in the CP paradigm, yet they are very good candidates to deal with and reason about uncertainty in the probabilistic paradigm, especially when the data is shaping an unknown distribution. The concept of p-boxes relies on the probabilistic approach that ranks probability distributions based on their stochastic dominance. It is a safe envelopment of the data whereabouts especially when it follows an unknown distribution. The {\em cdf} was selected due to its aggregated nature which enables the propagation of the information to the interval bounds in addition to its capability of easily ranking probability distributions within a p-box domain.

In Section \ref{sec:datarepresentation}, we have demonstrated that the p-box {\em cdf}-interval algebraic structure adds up quantitative information to real intervals which are adopted by convex models. We have also shown that the novel interval domain prevents probabilistic approximations which are carried on by models adopting possibilistic and probabilitic approaches. In Section \ref{sec:constraintreasoning}, we have shown that p-box {\em cdf}-interval operations adopt real-interval computations which are then projected linearly in the {\em cdf} domain. These operations guarrantee the envelopment of tuple computations exerted by each and every probability pair distributions lying within the intervals in the constraint relation. Moreover, the violation of the {\em cdf} ordering property shrinks the interval domain. Hence the realized solution space can be further pruned from the domain of real quantiles. The added value provided by the p-box {\em cdf}-intervals algebraic structure is a safe enclosure that bounds the data along with its whereabouts. This envelopment achieves tighter bounds on the output solution sets as opposed to those realized by convex models. In Section \ref{sec:empiricalevaluation}, we have evaluated the different modeling approaches, in terms of expressiveness and tractability, on a case study: an inventory management problem. We have shown how the p-box {\em cdf}-intervals intuitively envelop the uncertain data found in different modeling aspects with minimum overhead. 

In practice and based on our findings, stochastic CPs and probabilistic models are the slowest. Fuzzy models proved to have a better time performance and their output solutions are characterized to be reliable, i.e. they seek the satisfaction of all possible realizations. Convex models and the p-box {\em cdf}-intervals encapsulate all possible distributions of the solution set in a convex representation. The p-box {\em cdf}-intervals framework provides a range of quantities to order and a range of costs for each decision along with bounds on their data whereabouts.

The introduction of a novel framework to reason about data coupled with uncertainty due to ignorance or based on variability, paves the way to many fruitful research directions. We can list many in: studying models having variables following dependent probability distributions, exploring different search techniques, revisiting the framework within a dynamically changing environment, generalizing the framework to deal with all types of uncertainty by considering together vagueness and dynamicity, and last but not least applying the model to a variety of large scale optimization problems which target real-life engineering and management applications. 
\bibliographystyle{acmtrans}
\bibliography{pboxcdfintervals}
\label{lastpage}
\end{document}